\documentclass[aps,prb, preprint, citeautoscript, showpacs,floatfix,superscriptaddress]{revtex4-2}
\usepackage{bm}
\usepackage{amssymb}
\usepackage{graphicx}
\usepackage{amsmath}
\usepackage{textcomp}
\usepackage{colordvi}
\usepackage{multirow}
\usepackage{ulem}

\usepackage{color}
\usepackage{booktabs}

\begin{document}
\title{Theory of Optical Activity in Doped Systems with Application to Twisted Bilayer Graphene}
\author{K. Chang}
\email{knchang@ciomp.ac.cn}
\affiliation{GPL Photonics Lab, State Key Laboratory of Applied Optics, Changchun Institute of Optics, Fine Mechanics and Physics, Chinese Academy of Sciences, Changchun 130033, China.} 

\author{Z. Zheng}
\affiliation{GPL Photonics Lab, State Key Laboratory of Applied Optics, Changchun Institute of Optics, Fine Mechanics and Physics, Chinese Academy of Sciences, Changchun 130033, China.} 
\affiliation{University of Chinese Academy of Science, Beijing 100039, China.} 

\author{J. E. Sipe}
\affiliation{Department of Physics, University of Toronto, Toronto, Ontario M5S 1A7, Canada} 

\author{J. L. Cheng}
\email{jlcheng@ciomp.ac.cn}
\affiliation{GPL Photonics Lab, State Key Laboratory of Applied Optics, Changchun Institute of Optics, Fine Mechanics and Physics, Chinese Academy of Sciences, Changchun 130033, China.} 
\affiliation{University of Chinese Academy of Science, Beijing 100039, China.}
\date{\today}
\begin{abstract}

We theoretically study the optical activity in a doped system and derive the optical activity tensor from a light wavevector-dependent linear optical conductivity.
Although the light-matter interaction is introduced through the velocity gauge from a minimal coupling Hamiltonian, we find that the well-known ``false divergences'' problem can be avoided in practice if the electronic states are described by a finite band effective Hamiltonian, such as a few-band tight-binding model. 
The expression we obtain for the optical activity tensor is in good numerical agreement with a recent theory derived for an undoped topologically trivial gapped system.
We apply our theory to the optical activity of a gated twisted bilayer graphene, with a detailed discussion of the dependence of the results on twist angle, chemical potential, gate voltage, and location of rotation center forming the twisted bilayer graphene.

\end{abstract}
\maketitle

\section{Introduction}
\label{introduction}

Optical activity, also known as optical rotatory power, describes the rotation of the polarization direction as light propagates through an optically active medium \cite{Mahon2020,Rerat2021}. 
This phenomenon arises from the different responses to left and right circularly polarized light; the difference in absorption of the two polarizations is referred to as circular dichroism \cite{Ohnoutek2021}. 

Despite the broad application of circular dichroism in detecting the chirality of molecules \cite{Beaulieu2018,Ahn2020}, the relevant research in crystals is limited \cite{Tsirkin2018,Guo2022}. 
The quantum treatment of optical activity tensor is usually obtained 
from the charge-current density response \cite{Mahon2020,Zhong1993,Malashevich2010,Joensson1996} 
with the light-matter interaction included via the minimal coupling Hamiltonian \cite{Malashevich2010}.
However, this method can lead to ``false divergences'' when
the set of bands involved in the calculation is inevitably truncated
\cite{Sipe1993,Mahon2019}, and appropriate ``sum rules''  must be applied to show that the prefactors multiplying the divergent terms in fact vanish. 
To avoid this difficulty in calculating the optical activity, Mahon and Sipe \cite{Mahon2020} proposed a multipole moment expansion method for optical conductivity, where the macroscopic fields are introduced through the interactions with electric dipole, magnetic dipole, and electric quadrupole moments associated with Wannier functions at the lattice sites.
Despite the link this approach identifies with treating the optical response of isolated molecules, the derivation is complicated and the treatment is at present limited to undoped, topologically trivial insulators.

Alternate derivations --- particularly if they are simpler --- can often stimulate new research.
Starting from a minimal coupling Hamiltonian, in this paper we derive the expressions for the optical activity from the linear conductivity  tensor at finite light wavevector.
We then apply this method to study the optical activity of twisted bilayer graphene (TBG) using a simple tight-binding model \cite{Morell2017,Trambly2012,Morell2010} to describe its electronic states.
The optical response of TBG has been extensively studied \cite{Trambly2010,Andrei2020,Zheng2022,Tepliakov2020}, with the optical activity of undoped TBG investigated both experimentally \cite{Kim2016} and theoretically \cite{Morell2017}. 
In the undoped limit we find agreement with the results of Mahon and Sipe \cite{Mahon2020}, except for an extra term that indeed seems to exhibit a ``false  divergence''; however, while we cannot confirm analytically that the prefactor of this term vanishes, we can verify that it has a negligible value in our finite band tight-binding model.
And our approach is more general than that of Mahon and Sipe \cite{Mahon2020} in that it can be extended to doped systems; as well, it does not explicitly involve Wannier functions and the topological considerations necessary to construct them as localized functions. With our results in hand, we  explore the dependence of the optical activity tensor on twist angle, chemical potential, gate voltage, and location of rotation center forming twisted bilayer graphene.


\section{Models}
\label{methods}

\subsection{Optical activity tensor}
\label{meth-1}
An optically active material has different responses to left and right circularly polarized light, which are described by its linear optical conductivity $\sigma^{da}(\bm q,\omega)$. 
For an electric field 
${\bm E}({\bm r}, t) = {\bm E}({\bm q},\omega) e^{i({\bm q}\cdot{\bm r}-\omega t)} + {\rm c.c.}$,
the induced optical current is $\bm J(\bm r,t)=\bm J(\bm q,\omega)
e^{i({\bm q}\cdot{\bm r}-\omega t)} + {\rm c.c.}$ with 
$J^{d}({\bm q},\omega) = \sigma^{da}({\bm q},\omega) E^a({\bm q},\omega)$.
The Roman letters in the superscript denote Cartesian directions $x/y/z$, and the repeated superscripts are summed over. 
Without losing generality, considering an incident light propagating along the $z$-direction, the response to the circularly polarized light can be written as \cite{Cheng2017}
\begin{align}
  J^{\delta_1}(q\hat{\bm z},\omega) =  \sigma^{\delta_1\delta_2}(q\hat{\bm z},\omega)E^{\delta_2}(q\hat{\bm z},\omega)\,,
\end{align}
where $J^{\delta} =(J^x+i\delta J^y)/\sqrt{2}$ 
with $\delta=\pm$ gives the left ($+$)/right ($-$) circular component of the current; a similar definition applies to $E^{\delta}$. 
Then the diagonal response coefficients are 
\begin{align}
  \sigma^{\delta\delta}(q\hat{\bm z},\omega)
  &= \frac{\sigma^{xx}+\sigma^{yy}
    -i\delta (\sigma^{xy}-\sigma^{yx})}{2}\,.  
\end{align}
For nonzero $\sigma^{xy}-\sigma^{yx}$, the responses to the left and right circularly polarized lights are not the same, and the circular dichroism can be characterized by the ellipticity spectra
$\Psi=(\alpha_+-\alpha_-)/(2(\alpha_++\alpha_-))$, 
where $\alpha_\delta$  is the absorption of the $\delta$ circularly polarized light \cite{Li2013}. 
For two-dimensional materials, the absorption is proportional to $\text{Re}[\sigma^{\delta\delta}]$, and we can get
\begin{align}
\label{ell}
  \Psi = \frac{\text{Im}[\sigma^{xy}-\sigma^{yx}]}{2 \text{Re}[\sigma^{xx}+\sigma^{yy}]}\,.
\end{align}

The wave vector $\bm q$ appearing in the linear conductivity $\sigma^{da}(\bm q,\omega)$ is very small compared to most electron wave vectors, and up to the first order in $\bm q$ the conductivity can be expanded as 
\begin{align}
\label{sigma-ex}
\sigma^{da}({\bm q},\omega)=\sigma^{da}(\omega)+S^{dac}(\omega)q^c+\cdots \,,
\end{align} 
with $\sigma^{da}(\omega)\equiv\sigma^{da}({\bm 0}, \omega)$ giving the long wavelength limit, and 
\begin{align}
\label{eq:S}
\left.S^{dac}(\omega)\equiv
\frac{\partial\sigma^{da}({\bm q},\omega)}{\partial q^c}
\right|_{{\bm q}=0}\,.
\end{align}
The latter arises from effects of magnetic dipole and  electric
quadrupole, and modified
electric dipole effects \cite{Cheng2017}.
From macroscopic optics, it gives the response of the current to the spatial derivative of the electric field, including contributions following from Faraday's law.

For non-magnetic materials, the conductivity
tensor $\sigma^{da}(\omega)$ has either no off-diagonal components or
equal off-diagonal components
$\sigma^{da}(\omega)=\sigma^{ad}(\omega)$ \cite{boyd2020nonlinear}, and the optical activity mainly comes from the terms involving $S^{dac}(\omega)$, which is our focus in this paper. 
Because $S^{dac}(\omega)$ is a third-order tensor, it is nonzero only for crystals breaking inversion symmetry; 
more specifically, the nonzero $\text{Im}[\sigma^{xy}-\sigma^{yx}]$
indicates a chiral structure. 

\subsection{A microscopic response theory for $S^{dac}(\omega)$}
\label{meth-2}
For very weak electromagnetic fields, the Hamiltonian with the inclusion of light-matter interaction is taken from the minimal coupling as 
 %

\begin{align}
   \hat{H}(t) =
& \hat{H}_0 + \frac{-e}{2}\left[\hat{v}^a A^a({\bm r},t)+A^a({\bm r},t)\hat{v}^a\right] 
+ \frac{(-e)^2}{4} \left[\hat{M}^{da}A^d({\bm r},t)A^a({\bm r},t)\right.\notag\\
& \left.+A^d({\bm r},t)A^a({\bm r},t) \hat{M}^{da}\right]+\cdots\,,\label{eq:h}
\end{align}
with the electronic charge $e$,
$ \hat{\bm v}^a=[\hat{r}^a, \hat{H}_0]/(i\hbar)$, and
$\hat{M}^{da}=[\hat{r}^d, \hat{v}^a]/(i\hbar)$. Here $\hat{H}_0$ is the crystal Hamiltonian without external field, 
$\hat{v}^a$ is the velocity operator,
$\hat{M}^{da}$ is an operator associated with the mass term, 
and $A^a(\hat{\bm r},t)$ is the $a$th component of the vector
potential of the electromagnetic field.  The detailed
explanation of this Hamiltonian is listed in Appendix~\ref{app:H}. 
The field operator can be expanded as 
\begin{align}
  \hat \Psi(\bm r,t) &= \sum_n\int d\bm k \hat a_{n\bm k}(t) \phi_{n\bm k}(\bm r)\,.
\end{align}
Here $\phi_{n\bm k}(\bm r)=1/\sqrt{(2\pi)^3}e^{i\bm k\cdot\bm r}u_{n\bm k}(\bm r)$  are the band eigenstates of $\hat{H}_0$ with band eigenenergies $\varepsilon_{n\bm k}$ and periodic functions $u_{n\bm k}(\bm r)$,
and $\hat a_{n\bm k}(t)$ is an annihilation operator of this state.
The second quantization form of the Hamiltonian $\hat H(t)$ is 

\begin{align}
\hat H(t) = &\sum_{n} \int_{\text{BZ}} d\bm k \varepsilon_{n\bm k}\hat a_{n\bm k}^\dag \hat a_{n\bm k} + \int\frac{d\bm q}{(2\pi)^3}(-e) A^a_{\bm q}(t) \sum_{nm} \int_{\text{BZ}}d\bm k {\cal V}_{n\bm k+\bm q,m\bm k}^{(1);a}  \hat a_{n\bm k+\bm q}^\dag  \hat a_{m\bm k} \notag\\
&+\frac{1}{2}\int\frac{d\bm q_1d\bm q_2}{(2\pi)^6}(-e)^2A^a_{\bm q_1}(t)A^b_{\bm q_2}(t) 
\sum_{nm} \int_{\text{BZ}} d\bm k {\cal V}_{n\bm k+\bm q_1+\bm q_2,m\bm k}^{(2):ab}
    \hat  a_{n\bm k+\bm q_1+\bm q_2}^\dag  \hat a_{m\bm k} \,.
    \label{eq:h2}
\end{align}
Here $A^a_{\bm q}(t)=\int d\bm r A^a(\bm r,t)e^{-i\bm q\cdot\bm r}$ is the Fourier component of the vector potential, and the matrix elements are given by
\begin{align}
{\cal V}^{(1);a}_{n\bm k,m\bm k_1} &= \frac{1}{2}\sum_l\left( v^a_{nl\bm k} {\cal U}_{l\bm k,m\bm k_1} + {\cal U}_{n\bm k,l\bm k_1}v^a_{lm\bm k_1}\right) \,,\\
{\cal V}^{(2);ab}_{n\bm k,m\bm k_1} &= \frac{1}{2}\sum_l\left( M^{ab}_{nl\bm k} {\cal U}_{l\bm k,m\bm k_1} + {\cal U}_{n\bm k,l\bm k_1}M^{ab}_{lm\bm k_1}\right)\,,\\
  {\cal U}_{n\bm k,m\bm k_1}&=\int_{uc} \frac{d\bm r}{\Omega} u^*_{n\bm k}(\bm r) u_{m\bm k_1}(\bm r) \,,
\end{align}
where $v_{nm\bm k}^a$ and $M_{nm\bm k}^{ab}$ are the matrix elements of single particle operators $\hat{v}^a$ and $\hat{M}^{ab}$, respectively; $\Omega$ is the volume of the unit cell. 
The current density operator is given as 
\begin{align}
  \hat J^a({\bm q}, t) =& \int d\bm r \hat J^a(\bm r,t)e^{i\bm
                          q\cdot\bm r} = -\int d\bm r \frac{\delta\hat
                          H(t)}{\delta A^a(\bm r,t)}e^{i\bm
                          q\cdot\bm r} = -(2\pi)^3\frac{\delta \hat H(t)}{\delta A^a_{-\bm q}(t)} \,,
\end{align}
and the conductivity can be obtained from the Kubo formula as \cite{Mahan2000}
\begin{align}
\sigma^{da}(\bm q,\omega)=
& -g_s \frac{|e|^2}{i\omega} \sum_{nm}\int_{\text{BZ}}\frac{d\bm k}{(2\pi)^3} 
\frac{{\cal V}^{(1);d}_{m\bm k,n\bm k+\bm q} {\cal V}_{n\bm k+\bm q,m\bm k}^{(1);a} (f_{m\bm k}-f_{n\bm k+\bm q})}{\hbar\omega  -(\varepsilon_{n\bm k+\bm q}-\varepsilon_{m\bm k})+i0^+}
\notag\\
& -g_s\frac{|e|^2}{i\omega} \sum_n\int_{\text{BZ}}\frac{d\bm k}{(2\pi)^3} {\cal V}^{(2);da}_{n\bm k,n\bm k}f_{n\bm k}\,,
\label{eq:cond}
\end{align}
where the prefactor $g_s$ is the spin degeneracy
and $f_{mk}=\left[1+e^{(\varepsilon_{m\bm k}-\mu)/k_BT}\right]^{-1}$
gives the Fermi-Dirac distribution at the state
$\phi_{m\bm k}(\bm r)$ for a given temperature $T$ and
chemical potential $\mu$. 
Furthermore, the optical activity tensor in Eq.\,(\ref{eq:S}) is obtained as
\begin{align}
\label{oac}
S^{dac}(\omega) =& -g_s\frac{|e|^2}{2i\omega} \sum_{nm}\int_{\text{BZ}}\frac{d\bm k}{(2\pi)^3}
\left\{ 
       \frac{i v_{mn\bm k}^d \{r_{\bm k}^c,v_{\bm k}^a\}_{nm}- i\{r_{\bm k}^c, v_{\bm k}^d\}_{mn}v_{nm\bm k}^a}
       {\hbar\omega^+-\hbar\omega_{nm\bm k}} f_{mn\bm k} 
\right.  \notag\\
&\left. + \frac{v_{mn\bm k}^dv_{nm\bm k}^a}{\hbar\omega^+-\hbar\omega_{nm\bm k}}\left[\frac{\hbar (v_{nn\bm k}^c+v_{mm\bm k}^c)f_{mn\bm k}}{\hbar\omega^+-\hbar\omega_{nm\bm k}} - \frac{\partial(f_{n\bm k}+f_{m\bm k})}{\partial k^c}\right]\right\} \,,
\end{align}
with $\hbar\omega^+=\hbar\omega+i0^+$, $\hbar\omega_{nm\bm k}=\varepsilon_{n\bm k}-\varepsilon_{m\bm k}$ and $f_{nm\bm k}=f_{n\bm k}-f_{m\bm k}$.
Here, $\bm r_{nm\bm k}=(1-\delta_{nm})i\bm \nabla_{\bm q}U_{n\bm k,m\bm k+\bm q}|_{\bm q=0}$ 
gives the off-diagonal terms of the Berry connection, which can be
further calculated through  $\bm r_{nm\bm k}=\bm v_{nm\bm
  k}/(i\omega_{nm\bm k})$. The detailed derivation is given in Appendix~\ref{app:S}.

\subsection{Discussion of the expression Eq.\,(\ref{oac})}
\label{meth-3}
When the light-matter interaction is described using the vector potential,  
sum rules can be required to remove the ``false divergences''
occurring at zero frequency \cite{Sipe1993,Mahon2019}. 
To address this, we first perform an analysis of the expression in Eq.\,(\ref{oac}) for the behavior as $\omega\to0$. 
Utilizing
    \begin{subequations}
    \label{eq:exp1}
    \begin{align}
      \frac{1}{\omega^+-\omega_{nm\bm k}} =&
      -\frac{1}{\omega_{nm\bm k}}
      -\frac{\omega^+}{(\omega_{nm\bm k})^2}
      +\frac{({\omega^+})^2}{(\omega_{nm\bm k})^2(\omega^+-\omega_{nm\bm k})}\,,\\
      \frac{1}{(\omega^+-\omega_{nm\bm k})^2} =&
      \frac{1}{\omega_{nm\bm k}^2}+\frac{2\omega^+}{(\omega_{nm\bm k})^3}+\frac{(3\omega_{nm\bm k}-2\omega^+){(\omega^+)}^2}{(\omega_{nm\bm k})^3(\omega^+-\omega_{nm\bm k})^2}\,,
    \end{align}
  \end{subequations}
the expression of  $S^{dab}(\omega)$ is reorganized as 
\begin{align}
\label{Sadc}
  S^{dac}(\omega) = S_{f}^{dac}(\omega) + \frac{S_{-1}^{dac}}{\hbar\omega} + S_0^{dac} +
 S_r^{dac}(\omega)\,.
\end{align}
Here $S_f^{dac}(\omega)$ is the term involving derivatives of the populations;
$S_{-1}^{dac}$ and $S_0^{dac}$ are independent of $\omega$, they come from the first two terms of Eqs.~(\ref{eq:exp1}), respectively;
$S_r^{dac}(\omega)$ collects  all the remaining terms. 
Their expressions are given in Appendix~\ref{app:sterms}.

For an insulator,  the term $S_f^{dac}(\omega)$ vanishes, while $S_{r}^{dac}(\omega)$ is at least proportional to $\omega$, and is exactly the same as the optical activity tensor derived by Mahon and Sipe \cite{Mahon2020}. 
The term $S_{0}^{dac}$ vanishes for a system with time reversal symmetry, as we show in Appendix \ref{app:sterms}. 
Thus for a topologically trivial insulator the only difference between our result and that of Mahon and Sipe \cite{Mahon2020} is the term involving $S_{-1}^{dac}$. It leads to an $\omega^{-1}$ dependence in the conductivity,
and for a cold clean insulator,
a nonzero $S_{-1}^{dac}$ would lead to a divergent response as $\omega\to0$, which is unphysical; and thus a zero value of $S_{-1}^{dac}$ is required in this case. 
Adopting a nonzero value for $S_{-1}^{dac}$ would lead to a ``false divergence'' in the optical activity tensor, similar to those that arise in calculations of the linear susceptibility \cite{Sipe1993,Mahon2019}; as is the situation there, the vanishing of $S_{-1}^{dac}$ should be confirmed by the use of sum rules. 
Though we have not found an analytic proof that $S_{-1}^{dac}$ vanishes, as it must if our results are both to agree with Mahon and Sipe \cite{Mahon2020} for a cold, clean, topologically trivial insulator, and indeed to be physically meaningful for such a system as $\omega\rightarrow0$,  
we do numerically verify its value is negligible as long as a proper model Hamiltonian is adopted, as shown in the next section.

Compared to the approach of Mahon and Sipe \cite{Mahon2020}, ours has several advantages: 
(1) Since it does not rely on a Wannier function basis, and the topological considerations that must be invoked to identify when localized Wannier functions can be introduced, it should be free of any assumptions on the topology of the band structure.
(2) The contributions from free carriers are included. 
(3) Even beyond the small wavevector approximation --- for example, when the material interacts with confined light in nanostructures --- our expression in Eq.\,(\ref{eq:cond}) can still be applied. 
For a model Hamiltonian composed of finite numbers of bands, the correct use
of this expression for a numerical calculation  requires  the inclusion of all bands inside the model Hamiltonian and the integration over the whole
Brillouin zone (BZ).
Therefore, the direct application of the {\it ab initio} calculation
is not suitable, while the combination with {\it Wannier90} package
\cite{Pizzi2020}
could provide a finite band Hamiltonian.

\section{Results}
\label{results}

We apply the theory developed above to commensurate TBG, with the
electronic states described by  tight-binding models, as listed in
Appendix \ref{app:tb}. To use the expressions for a two dimensional
structure, the integration $\int_{\text{BZ}}\frac{d\bm k}{(2\pi)^3}$
 should be replaced  by $\int_{\text{BZ}}\frac{d\bm k}{(2\pi)^2}$ with a two
 dimensional $\bm k$. 
A TBG can be identified by the quantities $(m, n, \Delta, V_g, \bm \delta)$: the integer pairs $(m,n)$ are utilized to indicate the supercell and the twist angle $\theta$ \cite{Moon2013,Trambly2010};
$\Delta$ represents the on-site energy differences between A and B sites which can be induced by asymmetric substrate effects \cite{McCann2006,Cappelluti2012};
$V_g$ describes the potential differences between layers, which can be tuned by the gate voltage \cite{Nicol2008,Brun2015};
and $\bm\delta$ describes  the position of the rotation center, either at one carbon atom (for $\bm\delta=\bm 0$) or at the center of one hexagon (for $\bm\delta=\bm\delta_c\equiv 2(\bm a_1+\bm a_2)/3$).

In this work, we are interested in the TBGs with integer pairs 
$(m,n)=$ (2,1), (3,2), (4,3), (5,4), (6,5), (7,6), which correspond to twist angles of $21.8^\circ$, $13.2^\circ$, $9.4^\circ$, $7.3^\circ$, $6.0^\circ$, and $5.1^\circ$, respectively. 
For numerically evaluating the expressions in Eq.\,(\ref{oac}), the BZ is divided into a uniform grid of 1000 $\times$ 1000 for $(m,n)=$ (2,1), (3,2), (4,3), (5,4) and 500 $\times$ 500 for $(m,n)=$(6,5), (7,6), the temperature is taken as zero, and to avoid the divergences all frequencies $\omega$ are broadened as $\omega+i\gamma$ by a phenomenological parameter $\gamma$ with $\hbar\gamma=10$ meV, unless otherwise specified. 
The derivative of the population is converted to
\begin{align}
\frac{\partial f_{n\bm k}}{\partial k^a} &= -\hbar v_{nn\bm k}^a \frac{\partial f_{n\bm k}}{\partial \mu}\,,
\end{align}
and ${\partial f_{n\bm k}}/{\partial \mu}$ is numerically  calculated by a difference between chemical potentials $\mu+\delta\mu/2$ and $\mu-\delta\mu/2$ with $\delta\mu=5$ meV. 

Before presenting numerical results, we perform a symmetry analysis on the tensor components. 
The crystal symmetry depends on the values of $V_g$ and $\bm\delta$. 
For $V_g=0$, the point group is $D_6$ for $\bm\delta=\bm\delta_c$ and $D_3$ for $\bm\delta=\bm 0$, and they are reduced to $C_6$ and $C_3$ for nonzero $V_g$, respectively.  
The nonzero components of the optical activity tensor are summarized
in Table\,\ref{table},  and all cases include the tensor components $xyz=-yxz$, $xzy=-yzx$ and $zxy=-zyx$.
Considering the geometry discussed in Section \ref{meth-1}, we mainly
focused on the component $S^{xyz}$.
\begin{table*}[htp]
	\centering
	\setlength{\abovecaptionskip}{0.4cm}
	\setlength{\belowcaptionskip}{0.3cm}
	\small
	\caption{Nonzero components of the optical activity tensor for
          different parameters \cite{boyd2020nonlinear}.}
	\vspace{3pt}
	\setlength{\tabcolsep}{2.5mm}
	{	
	\begin{tabular}{ccl}
		\hline\hline		
TBG Parameters & Point Group & Nonzero Tensor Components\\
	    \hline
$(m, n, 0, 0, \bm 0)$ & $D_3$ & $xxx=-xyy=-yyx=-yxy$, $xyz=-yxz$, \\
	                          &       & $xzy=-yzx$, $zxy=-zyx$ \\
	    \hline
$(m, n, 0, V_g, \bm 0)$  & $C_3$ & $xxx=-xyy=-yyz=-yxy$, $xyz=-yxz$,  \\
                         &       & $xzy=-yzx$, $zxy=-zyx$, $xzx=yzy$, $xxz=yyz$, \\
                         &       & $zxx=zyy$, $yyy=-yxx=-xxy=-xyx$,  $zzz$ \\
	    \hline
$(m, n, 0, 0, \bm \delta_c)$ & $D_6$ & $xyz=-yxz$, $xzy=-yzx$, $zxy=-zyx$ \\
	    \hline
$(m, n, 0, V_g, \bm \delta_c)$ & $C_6$ & $xyz=-yxz$, $xzy=-yzx$, $zxy=-zyx$, \\
	                           &       & $xzx=yzy$, $xxz=yyz$, $zxx=zyy$, $zzz$ \\
	    \hline\hline
	\end{tabular}}
	\label{table}
\end{table*}

\subsection{Comparison with literature}

\begin{figure}[htp]
 \centering
		\includegraphics[scale=1.]{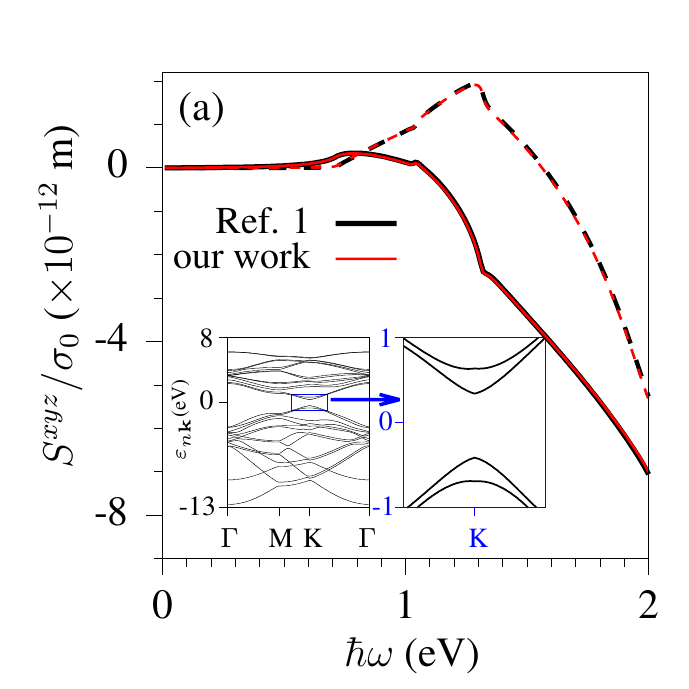}
		\includegraphics[scale=1.]{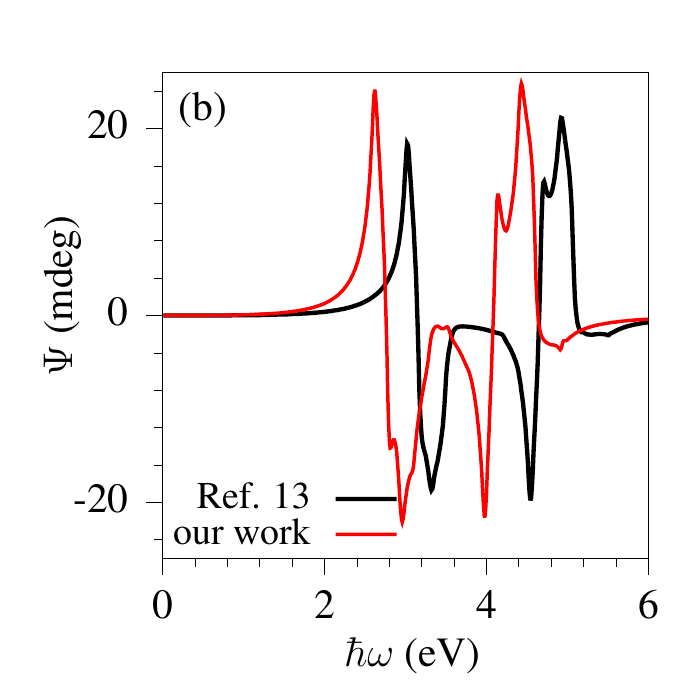}
	\caption{
Comparisons between our theory and those in literature. 
(a) The spectra of $S^{xyz}$ using the method in Ref.\,\citenum{Mahon2020} (black lines) and in our work (red lines) for a $(2,1,2\,{\rm eV},0,\bm\delta_c)$-TBG.  
The solid (dashed) curves give the real (imaginary) parts. 
The insets are the band structure of a gapped TBG. 
(b) The ellipticity spectra for a $(2,1,0,0,\bm\delta_c)$-TBG using the Hamiltonian provided in Ref.\,\citenum{Morell2017} (black lines) and in our work (red lines). 
The damping energy is adjusted to 25 meV for comparison.
}
	\label{gap}
\end{figure}

We numerically validate the derived expression by comparing with two theories presented in literature \cite{Mahon2020,Morell2017}. 
The first concerns the ``sum rules'' of $S_{-1}^{dab}=0$ 
in an undoped topologically trivial insulator at zero temperature,
which vanishes automatically for an insulator in the theory by Mahon and Sipe \cite{Mahon2020}.
To get a gapped system, we choose the parameters as $(m, n, \Delta, V_g, \bm\delta)=(2, 1, 2\,{\rm eV}, 0, \bm \delta_c)$, which generates a bandgap of 0.77 eV. 
In Fig.\,\ref{gap}\,(a), we show the spectra of $S^{xyz}$ obtained from our expressions and those from the expression by Mahon and Sipe \cite{Mahon2020}. 
The excellent agreement indicates the existence of the sum rule $S_{-1}^{dab}=0$. 
More specifically, we find
$|S_{-1}^{xyz}/\sigma_0|\approx 3.17\times10^{-30}$ ${\rm eV \cdot m}$
with  $\sigma_0=e^2/4\hbar$.
Note that such negligible values are obtained with all bands in the
model Hamiltonian and with the inclusion of whole BZ. 
When only half of the bands on both sides of the Fermi level are used in the calculation, for example, we obtain $|S_{-1}^{xyz}/\sigma_0|\approx 4.70\times10^{-17}$ ${\rm eV \cdot m}$. 
Similarly, it becomes $|S_{-1}^{xyz}/\sigma_0|\approx 4.43\times10^{-17}$ ${\rm eV \cdot m}$, 
if the calculation is performed for all bands but includes only k points satisfying that the transition energies between the lowest conduction band and the highest valence band are less than 2 eV.

Then we compare the ellipticity spectra obtained from our expressions with those of Morell {\it et al.}, where a different tight-binding model is used (also listed in Appendix~\ref{app:sterms}).
As shown in Fig.\,\ref{gap}\,(b), the results from two tight-binding models are very similar with respect to the spectra shape, except that the locations of the peaks and valleys are shifted, which is ascribed to the different tight-binding parameters.

\subsection{Twist angle dependence of $S^{dac}$}

\begin{figure}[htp]
 \centering
        \includegraphics[scale=1.]{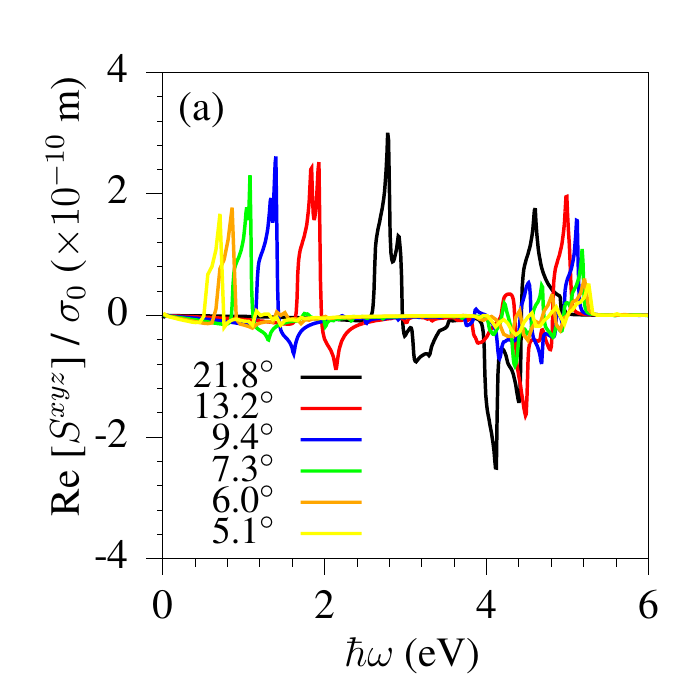}
        \includegraphics[scale=1.]{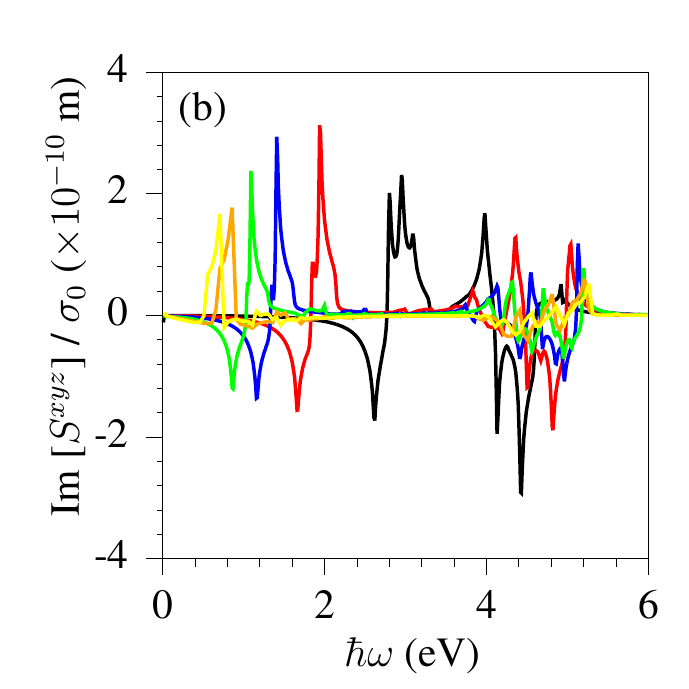}
	\caption{
The real (a) and imaginary (b) parts of $S^{xyz}$ for twist angles  21.8$^{\circ}$, 13.2$^{\circ}$, 9.4$^{\circ}$, 7.3$^{\circ}$, 6.0$^{\circ}$, and 5.1$^{\circ}$.}
	\label{size}
\end{figure}

Figure \ref{size}\,(a) and (b) show the spectra of $S^{xyz}$ for the $(m,n,0,0,\bm\delta_c)$-TBG at various $(m,n)$ or twist angles.
The spectra for each twist angle consist mainly of two peaks in different regions of photon energies: 
One is at low photon energies --- less than 3 eV --- which moves to smaller photon energies for smaller twist angles; 
the other is at the high-energy region around 4--5 eV, which changes  little with twist angles.
We find both features are associated with the optical transitions around Van Hove singularity  (VHS) points. 
In TBG, there are two types of VHS: (1) The first is formed by the intersection of the two Dirac
cones of the upper and lower graphene layer; it lies in the lowest
conduction band and highest valence band, and the transition energy
reduces with the decreasing twist angle \cite{Zheng2022}. 
(2) The second is inherited from the VHS of each monolayer graphene, for which
the energy changes little with twist angle. 
Because our approach requires the inclusion of all bands and the integration is over the whole BZ, the calculation becomes extremely time consuming for small twisted angles, especially for the ``magic angles'', due to the large number of bands.

\subsection{Chemical potential dependence of $S^{dac}$}

\begin{figure*}[htp]
 \centering
 	\includegraphics[scale=0.9]{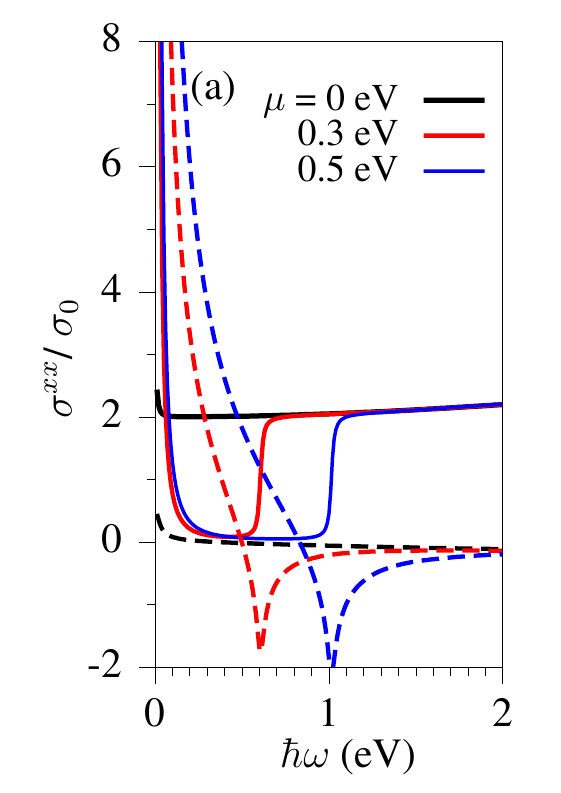}
	\includegraphics[scale=0.9]{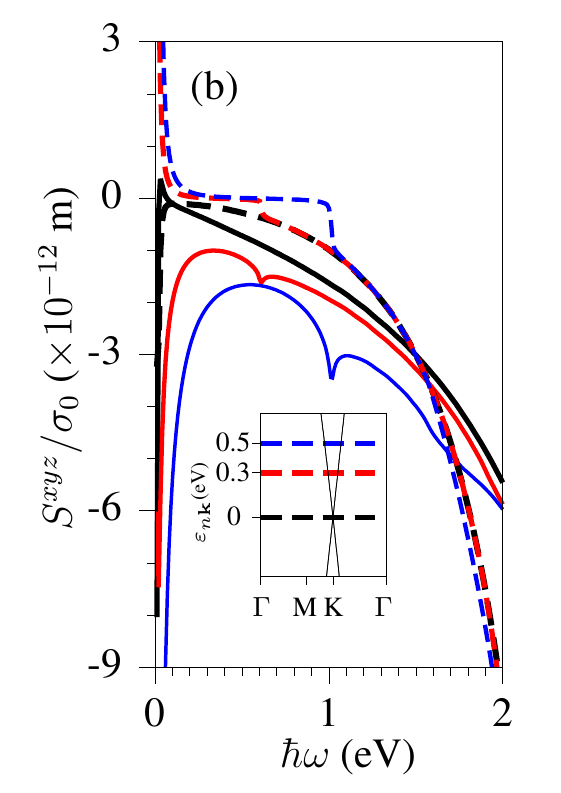}
	\includegraphics[scale=0.9]{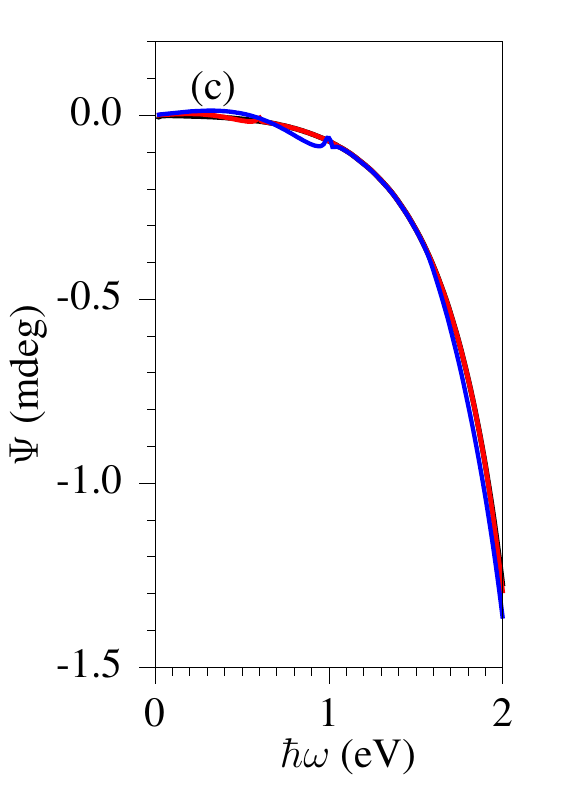}
	\caption{Spectra of (a) $\sigma^{xx}$, (b) $S^{xyz}$, and (c) $\Psi$ for the $(2,1,0,0,\bm\delta_c)$-TBG at different chemical potentials $\mu$ = 0, 0.3 and 0.5 eV. 
	The solid (dashed) curves give the real (imaginary) part.  
    The inset in (b) indicates the position of chemical potentials in the band structure.}
	\label{chem}
\end{figure*}
Next, we discuss the spectra of $S^{dac}$ for $(2,1,0,0,\bm\delta_c)$-TBG at different chemical potentials $\mu=0$, 0.3, and 0.5 eV.
Figure \ref{chem} displays the spectra of the linear conductivity
$\sigma^{xx}$, the optical activity $S^{xyz}$, and the ellipticity
$\Psi$ at the photon energies less than 2 eV, where the chemical
potential has significant effects. 
For undoped TBG, the real part of the linear conductivity has a flat curve with values around $2\sigma_0$ \cite{Tabert2013} and the imaginary part is near zero.
Due to the zero gap, the real part of $S^{xyz}$ is nearly proportional
to the photon energy, while the imaginary part shows a quadratic
relation; these tendencies are very similar to that of a gapped
insulator with the gap approaching zero.
As the chemical potential increases to 0.3 eV or 0.5 eV, there are
free carriers and partially filled bands, which lead to the appearance
of the intraband transition for states around the Fermi surface and the Pauli blocked interband transition for states below the Fermi level.
For the conductivity $\sigma^{xx}$, the intraband transition gives a
Drude contribution $\propto (\omega+i\gamma)^{-1}$ at small photon
energies; while for optical activity $S^{xyz}$, the intraband
transition lies in the term of $S_f$,  and it can give a Drude-like
contribution $\propto(\omega+i\gamma)^{-2}$; note that, due to free
carriers, the term $S_{-1}$ becomes nonzero and contributes by
$(\omega+i\gamma)^{-1}$.
Therefore, for a finite chemical potential, both $\sigma^{xx}$ and $S^{xyz}$
 show behavior that would be divergent for small photon energies.
The interband transition leads to an effective band gap $E_g\approx 2|\mu|$,
which determines the onset of the real part of $\sigma^{xx}$ and the
imaginary part of $S^{xyz}$. Similarly, the imaginary part of
$\sigma^{xx}$  and the real part of $S^{xyz}$ show divergent peaks in
their magnitudes as the photon energy matches the effective gap. 
Despite of the rich structure in
the spectra of $\sigma^{xx}$ and $S^{xyz}$, changing the chemical potential only slightly changes the value of $\Psi$, as shown in Fig.\,\ref{chem}\,(c).

\begin{figure}[htp]
    \centering
	\includegraphics[scale=1]{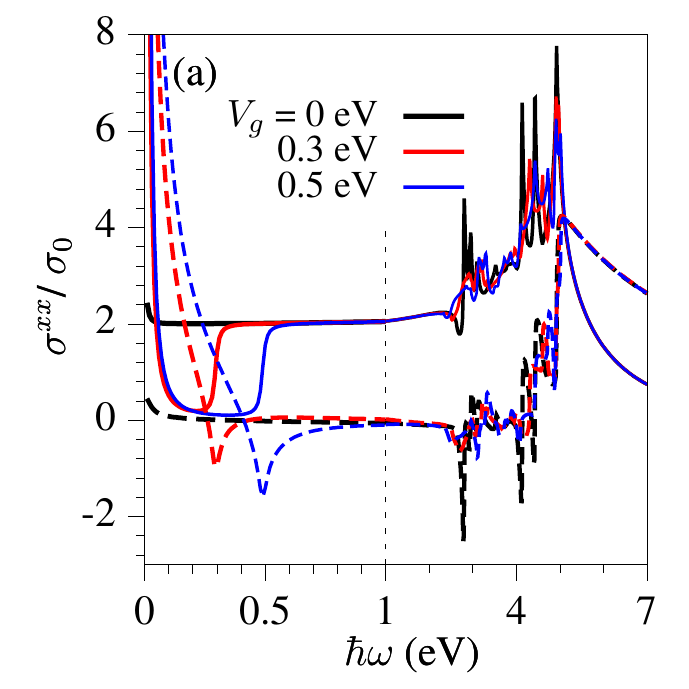}
	\includegraphics[scale=1]{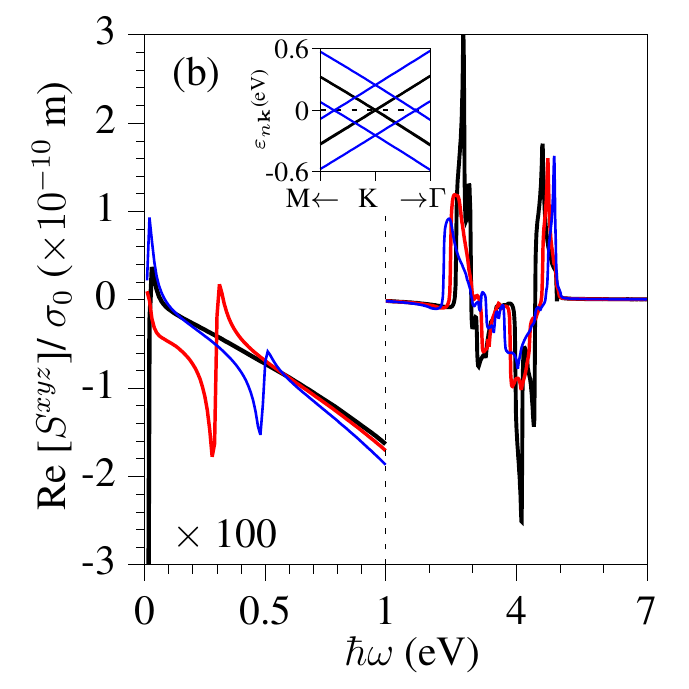} \\
	\includegraphics[scale=1]{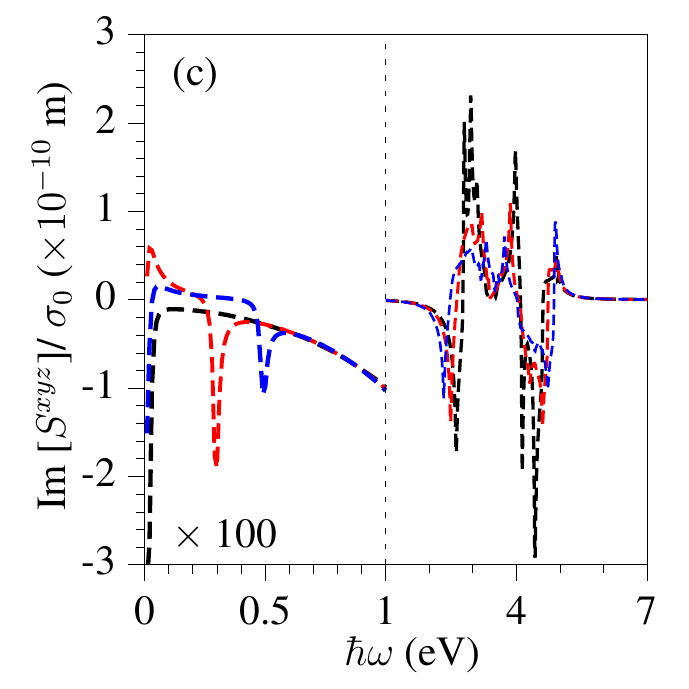}
    \includegraphics[scale=1]{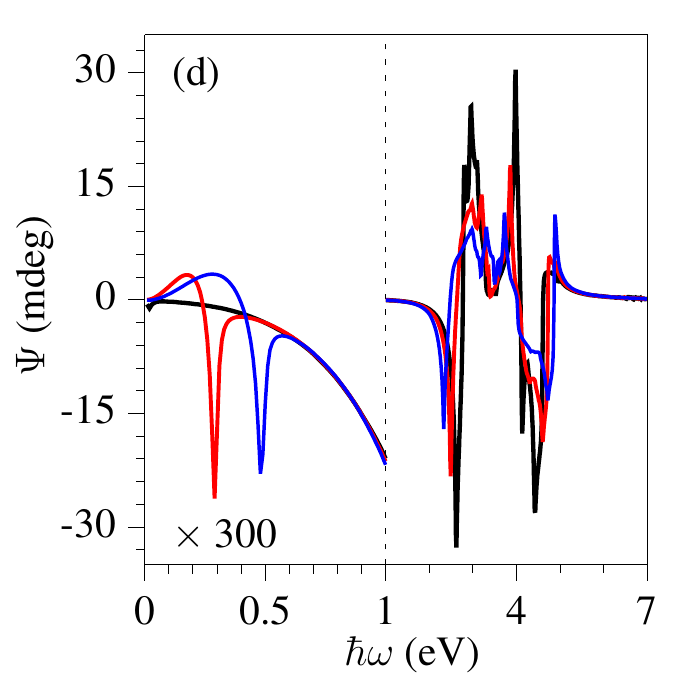}
	\caption{The spectra of (a) $\sigma^{xx}$, (b) real part of $S^{xyz}$, (c) imaginary part of $S^{xyz}$, and (d) $\Psi$
	of the $(2,1,0,V_g,\bm\delta_c)$-TBG for different gate voltages $V_g=$ 0, 0.3 and 0.5 eV.  
	Values at $\hbar\omega<1$ eV of $S^{xyz}$ and of $\Psi$ are zoomed in 100 and 300 times, respectively.
The inset in (b) illustrates how the gate voltage affects the band
structure around the Dirac points.}
	\label{gate}
\end{figure}
\subsection{Gate voltage dependence of $S^{dac}$}
We now turn to the effects of the gate voltage on the optical activity.
Figure \ref{gate} gives the spectra of linear conductivity $\sigma^{xx}$, optical activity $S^{xyz}$, and the related ellipticity spectra $\Psi$ of $(2,1,0,V_g,\bm\delta_c)$-TBG for different gate voltages  $V_g=$ 0, 0.3, 0.5 eV.  
Note that while the applied gate voltage lowers the symmetry and leads to new nonzero components of the optical activity tensor, we still focus on the component $S^{xyz}$ here. 
The gate voltage affects the band structure in two ways \cite{Nicol2008,Brun2015,Yu2019}: 
one is to split the near degenerate Dirac cones at the Dirac points by lifting the energy of one cone and lowering that of the other,  which is analogous to tuning the upper and lower graphene layers with different chemical potentials; 
and the other is to split both types of VHS points. 
Therefore, the real part of linear conductivity shows a dip-peak curve at photon energies around $V_g$, and its imaginary part shows a valley, due to the inhibited interband transition at the lower-energy region.
Here, the effects of the gate voltage on the linear conductivity are similar to those of the chemical potential.
As for the the optical activity, a dip-peak structure in the real part and a valley in the imaginary part are observed. 
However, at the higher energy region, unlike the chemical potential,
the gate voltage significantly lowers the peak values of both the linear conductivity and the optical activity.
As shown in Fig.\,\ref{gate}\,(d), the gate voltage has significant effects on the ellipticity spectra, where at the lower energy part a peak appears at energy around $V_g$, and at the higher energy part the peak values are reduced with the increasing $V_g$.

\subsection{Rotation center dependence of $S^{dac}$}
\begin{figure*}[htp]
 \centering
	\includegraphics[scale=0.9]{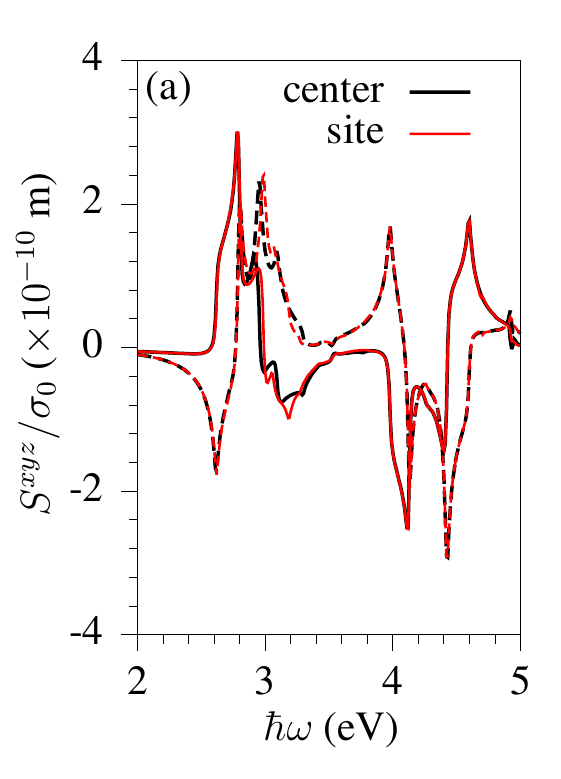}
	\includegraphics[scale=0.9]{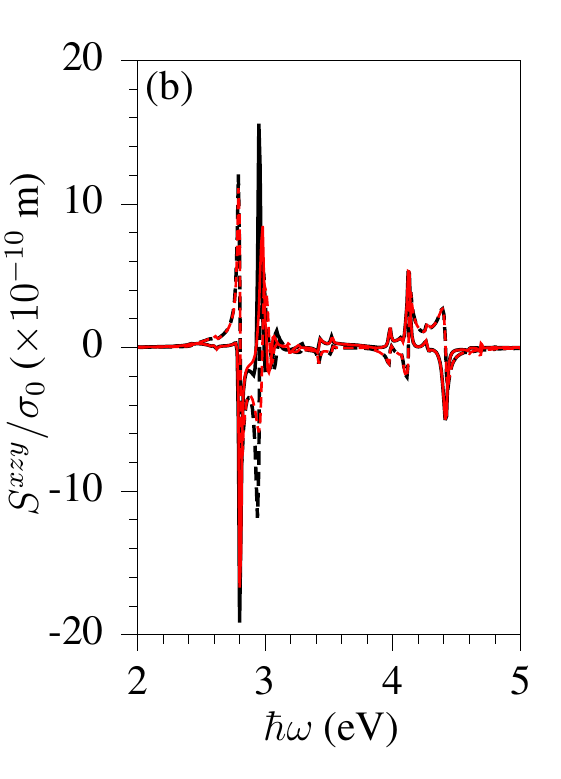} 
	\includegraphics[scale=0.9]{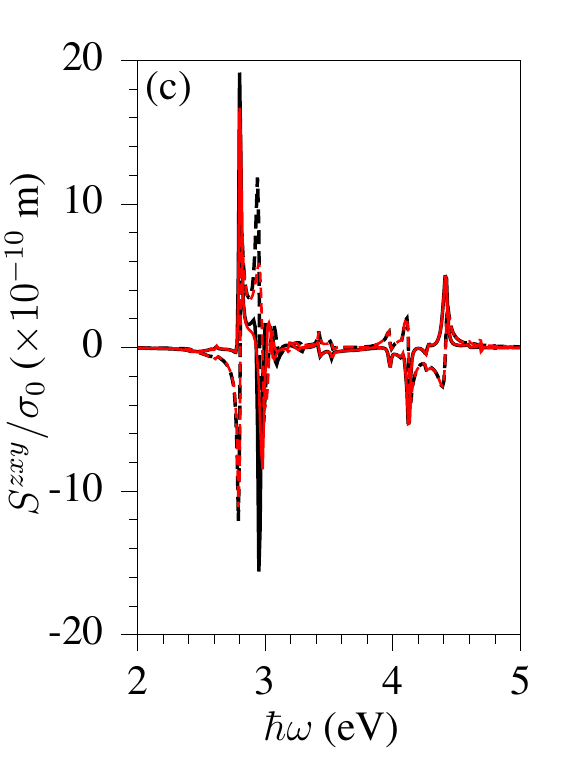}
	\caption{
	Spectra of optical activity tensors (a) $S^{xyz}$, (b) $S^{xzy}$, and (c) $S^{zxy}$ of TBG with rotation centers at the atom (red curves) and at the center of the hexagon (black curves).}
	\label{rota}
\end{figure*}

Here we consider how the rotation center affects the optical activity
tensor. 
Figure \ref{rota} gives the spectra of the three nonzero tensor components for the $(2,1,0,0,\bm\delta_c)$-TBG and $(2,1,0,0,\bm 0)$-TBG.
The results show minor differences. 
This is very similar to the effect of the rotation center on the linear conductivity \cite{Moon2013}, mostly because the rotation center modifies the band structure very slightly.
Although the symmetry analysis gives additional nonzero independent tensor components for $\bm \delta=\bm 0$, their values are about several orders of magnitude smaller than other nonzero components.
%

\section{Conclusions}
\label{conclusions}
%
We have derived the optical activity tensor by describing the light-matter interaction using the minimal coupling Hamiltonian.
Considering the ``false divergences'' that can arise in this framework, we found that the expression for the prefactor that should vanish is numerically stable and negligibly small if a finite band model Hamiltonian is used and the integration is extended over the whole Brillouin zone.  
Our expressions are valid  for doped systems as well as topological nontrivial materials.  
We then applied the expressions to a twisted bilayer graphene and studied the effects of the twist angle, chemical potential, gate voltage, and rotation center. 
The doping can cause a Drude-like contribution to the optical activity tensor for small photon energies, the imaginary parts of the tensor components show  onset from the interband transition around the chemical potential induced gap, and the real parts exhibit peaks at the same frequencies.
The chemical potential affects the optical activity at very high energies.
The gate voltage can modify the band structure at either low or high photon energies, and correspondingly affects the optical activity significantly. 
We find that the optical activity of twisted bilayer graphene is weakly affected by the rotation center.

Our theory provides a very preliminary study using the minimal coupling Hamiltonian in the single particle approximation. 
Considering that the excitonic effects are very important for the optical response of two-dimensional materials, it is necessary to extend the theory to include  excitonic effects as well as local field corrections \cite{Joensson1996}.

\acknowledgments
This work has been supported by Scientific research project of the
Chinese Academy of Sciences Grant No. QYZDB-SSW-SYS038, National
Natural Science Foundation of China Grant No. 12034003 and 12004379.
J.L.C. acknowledges support from Talent Program of CIOMP.
J.E.S. acknowledges support from the Natural Sciences and Engineering Research Council
of Canada.

%

\appendix
\section{Form of Eq.~(\ref{eq:h})\label{app:H}}
We briefly explain how to obtain the Hamiltonian in Eq.~(\ref{eq:h}). For an unperturbed Hamiltonian $\hat H_0(\bm r,\bm p)$, which could be used to describe the system with a static magnetic field as well, the interaction between the electrons and the light is introduced by a minimal coupling to a time dependent the vector potential $\bm A(\bm r,t)=\int\frac{d\bm q}{(2\pi)^3}\bm A_{\bm q}(t)e^{i\bm q\cdot\bm r}$ as  the Hamiltonian $\hat{H}_0(\bm r,\bm p-e\bm A(\bm r,t))$. 
After expanding it in terms of $\bm A_{\bm q}(t)$, we formally get
  \begin{align}
    \hat H_0(\bm r,\bm p-e\bm A(\bm r,t)) =& \hat H_0(\bm r,\bm p) 
    +(-e)\int\frac{d\bm q}{(2\pi)^3} \hat {\cal V}^{(1);a}_{\bm q}(\bm r,\bm p) A_{\bm q}^a(t) \notag\\
    &+ \frac{(-e)^2}{2}  \int\frac{d\bm q}{(2\pi)^3}
    \hat {\cal V}^{(2);ab}_{\bm q_1+\bm q_2}(\bm r,\bm p) A_{\bm
      q_1}^a(t) A_{\bm q_2}^b(t)+\cdots\,. \label{eq:exph}
  \end{align}
  For a standard local Hamiltonian $\hat H_0 (\bm r,\bm p)=\frac{\bm
    p^2}{2m_0}+V(\bm r)$, the expressions of $\hat {\cal V}^{(1)}$ and
  $\hat {\cal V}^{(2)}$ are
  \begin{align}
     \hat {\cal  V}^{(1);a}_{\bm q}(\bm r,\bm p)  &=\frac{1}{2}\left(
    \frac{\hat{\bm p}}{m_0}e^{i\bm q\cdot\bm r}+e^{i\bm q\cdot\bm
    r}\frac{\hat{\bm p}}{m_0}\right)\,,\\
    \hat {\cal  V}^{(2);ab}_{\bm q}(\bm r,\bm p)
                                                  &=\delta_{ab}\frac{1}{m_0}
                                                    e^{i\bm q\cdot\bm r}\,,
  \end{align}
  respectively.
  However,  in many effective models or \textit{ab initio} calculations, $\hat
  H_0(\bm r,\bm p)$ includes the contributions from
  nonlocal potentials, or perhaps even only its matrix elements between
tight-binding basis functions are specified.
In such situations, the explicit form of the light-matter
  interaction term is not very straightforward. To obtain a 
  Hamiltonian that is appropriate for our calculations, and especially to eliminate
  in practice the ``false divergences'' that can plague minimal coupling
calculations, we find it is necessary to choose the expansion coefficients so that gauge invariance is satisfied, at least up to the linear order of  light wave
  vector $\bm q$. 
  For a gauge transformation
  \begin{subequations} 
  \begin{align}
  \boldsymbol{A}(\bm r,t)&\rightarrow\boldsymbol{A}(\bm r,t)+\nabla g(\bm r,t)\,,\\
  \phi(\bm r,t)&\rightarrow\phi(\bm r,t)-\frac{\partial g(\bm r,t)}{\partial t}\,,
\end{align}
\end{subequations}
 where $\phi(\bm r,t)$ is the scalar potential
and $g(\bm{r})=g_{\bm{q}}e^{i\bm{q}\cdot\bm{r}}+{\rm c.c.}$
implements the gauge transformation, we require that the form (\ref{eq:exph}) of the coupling be the same whether the new or old potentials are employed. 
Using
  \begin{align}
    \hat H_0(\bm r,\bm p-e\bm A(\bm r,t) - e \bm\nabla g(\bm r)) =
    e^{i\frac{e}{\hbar} g(\bm r)} \hat H_0(\bm r,\bm p-e\bm A(\bm
    r,t))  e^{-i\frac{e}{\hbar} g(\bm r)} \,,
  \end{align}
and employing a perturbative expansion of the right-hand-side with respect to the
  orders of $g_{\bm q}$, gauge invariance requires that 
  \begin{align}
    iq^a \hat {\cal V}^{(1);a}_{\bm q}(\bm r,\bm p) &= \frac{1}{i\hbar}
                                                 [e^{i\bm q\cdot\bm
                                                 r},\hat H_0(\bm r,\bm
                                                 p)]\,,\label{eq:gv1}\\
    iq_1^a\hat {\cal V}^{(2);ab}_{\bm q_1+\bm q_2}(\bm r,\bm p) &=
                                                 \frac{1}{i\hbar}[e^{i\bm
                                                 q_1\cdot\bm r}, {\cal
                                                 V}^{(1);b}_{\bm
                                                 q_2}(\bm r,\bm
                                                             p)]\,. \label{eq:gv2}
  \end{align}
  It can be verified that the expressions 
  \begin{align}
    \hat {\cal  V}^{(1);a}_{\bm q}(\bm r,\bm p) &= \frac{1}{2}\left[\hat v^a(\bm r,\bm p)
    e^{i\bm q\cdot\bm r} + e^{i\bm q\cdot\bm r}\hat v^a(\bm r,\bm
                              p)\right]\,,\label{eq:calV1}\\
    \hat{\cal V}^{(2);ab}_{\bm q}(\bm r,\bm p) &= \frac{1}{2}\left[\hat
                                         M^{ab}(\bm r,\bm p)e^{i\bm
                                         q\cdot\bm r}+e^{i\bm
                                                 q\cdot\bm r}\hat
                                                 M^{ab}(\bm r,\bm p)\right]\,,\label{eq:calV2}
  \end{align}
  satisfy the conditions in Eqs.~(\ref{eq:gv1}, \ref{eq:gv2}) 
  up to the first order in $\bm q$. Substituting $\hat{\cal
    V}^{(1)}$ and $\hat{\cal V}^{(2)}$ in Eqs.~(\ref{eq:calV1}, \ref{eq:calV2})
   back to Eq.~(\ref{eq:exph}), we get Eq.~(\ref{eq:h}).  

\section{Derivation of Eq.~(\ref{oac})\label{app:S}}
For the conductivity in Eq.~(\ref{eq:cond}), only the first
term at the right hand side include $\bm q$. Using the Berry connection 
\begin{align}
  \bm\xi_{mn\bm k}&=i\bm\nabla_{\bm q}U_{m\bm k,n\bm k+\bm q}|_{\bm
                    q=\bm 0}\,,
\end{align}
and 
\begin{align}
  \left.\frac{\partial}{\partial q^c}{\cal V}^{(1);d}_{m\bm k,n\bm k+\bm
  q}\right|_{\bm q=0}
  &= \frac{1}{2}\frac{\partial v_{mn\bm k}^a}{\partial k^c}+\frac{i}{2}\sum_l
    \left[v_{ml\bm k}^ar_{lm\bm k}^c + r_{nl\bm k}^c v_{lm\bm
    k}^a\right] + \frac{i}{2}\left(\xi_{nn\bm k}^c+\xi_{mm\bm
    k}^c\right)v_{mn\bm k}^a\notag\\
  &=\frac{1}{2}\frac{\partial v_{mn\bm k}^a}{\partial k^c} +
    \frac{i}{2}\left\{r_{\bm k}^c,v_{\bm k}^a\right\}_{mn}+ \frac{i}{2}\left(\xi_{nn\bm k}^c+\xi_{mm\bm
    k}^c\right)v_{mn\bm k}^a\,,
\end{align}
where the off-diagonal terms of $\bm\xi_{nm\bm k}$ is noted as $\bm
r_{nm\bm k}$, we get
\begin{equation}
  \left.\frac{\partial}{\partial q^c}\left[{\cal V}^{(1);d}_{m\bm k,n\bm k+\bm
  q}{\cal V}^{(1);a}_{n\bm k+\bm
  q,m\bm k}\right]\right|_{\bm q=0}
=\frac{1}{2}\frac{\partial (v_{mn\bm k}^dv_{nm\bm k}^a)}{\partial
    k^c}+\frac{i}{2}v_{mn\bm k}^d\{r_{\bm k}^c,v_{\bm
    k}^a\}_{nm}-\frac{i}{2}\{r_{\bm k}^c,v_{\bm k}^d\}_{mn}v_{nm\bm k}^a\,,\label{eq:e1}
\end{equation}
with the notation $\{r_{\bm k}^c,v_{\bm k}^a\}_{nm} = \sum_l (r_{nl\bm k}^cv_{lm\bm k}^a+v_{nl\bm k}^ar_{lm\bm k}^c)$.

In calculating $S^{dac}$, the first term in Eq.~(\ref{eq:e1}) appears as
\begin{align}
 \propto \int_{\text{BZ}}\frac{d\bm k}{(2\pi)^3}\frac{ \frac{\partial (v_{mn\bm k}^dv_{nm\bm k}^a)}{\partial
    k^c}f_{mn\bm k}}{\hbar\omega^+-\hbar\omega_{nm\bm k}} = -\int_{\text{BZ}}\frac{d\bm k}{(2\pi)^3} (v_{mn\bm k}^dv_{nm\bm k}^a)\frac{\partial}{\partial
    k^c}\frac{ f_{mn\bm k}}{\hbar\omega^+-\hbar\omega_{nm\bm k}}\,.
\end{align}
Using all these expressions, we find Eq.~(\ref{oac}).

\section{Expressions of $S_f$, $S_{-1}$, $S_0$, and $S_r$ \label{app:sterms}}
Substituting Eqs.\,(\ref{eq:exp1}) into Eq.\,(\ref{oac}), we can
directly get
\begin{subequations}
  \begin{align}
  S^{dac}_f(\omega) &=g_s 
	\frac{|e|^2}{2i\omega} \sum_{nm}\int_{\text{BZ}}\frac{d\bm k}{(2\pi)^3}
	\frac{v_{mn\bm k}^dv_{nm\bm k}^a}{\hbar\omega^+-\hbar\omega_{nm\bm k}}
	\frac{\partial(f_{n\bm k}+f_{m\bm k})}{\partial k^c}\,,
  \end{align}
  \begin{align}
    S^{dac}_{-1}
    =&-g_s 
      \frac{|e|^2}{2i\hbar} \sum_{nm}\int_{\text{BZ}}
      \frac{d\bm k}{(2\pi)^3}\left[ r_{mn\bm k}^dr_{nm\bm k}^a(v_{nn\bm k}^c+v_{mm\bm k}^c)\right.\notag\\
    &\left. -( r_{mn\bm k}^d \{r_{\bm k}^c,v_{\bm k}^a\}_{nm}+
      \{r_{\bm k}^c, v_{\bm k}^d\}_{mn}r_{nm\bm k}^a) \right]f_{mn\bm k} \,,
  \end{align}
  \begin{align}
    S_0^{dac}    
    =&-g_s 
	\frac{|e|^2}{2i\hbar} \sum_{nm}\int_{\text{BZ}}
	\frac{d\bm k}{(2\pi)^3}\left[2 r_{mn\bm k}^dr_{nm\bm k}^a(v_{nn\bm k}^c+v_{mm\bm k}^c)\right.\notag\\
    &\left. -( r_{mn\bm k}^d \{r_{\bm k}^c,v_{\bm k}^a\}_{nm}+
      \{r_{\bm k}^c, v_{\bm k}^d\}_{mn}r_{nm\bm k}^a) \right]
      \frac{f_{mn\bm k}}{\omega_{nm\bm k}} \,,
  \end{align}
  and
  \begin{align}
    S^{dac}_r(\omega)
    &=g_s 
	\frac{e^2\hbar \omega }{2i} 
    \int_{\text{BZ}}\frac{d\bm k}{(2\pi)^3}\sum_{mn}
      \frac{X^{dac}_{mn\bm k}(\omega)f_{nm\bm k}}
      {\hbar\omega_{mn\bm k}(\hbar\omega_{mn\bm k}-\hbar\omega^+)}\,,
       \label{S-r}
  \end{align}
\end{subequations}
with
  \begin{align}
    X^{dac}_{mn\bm k}=&r_{nm\bm k}^d\{v^a_{\bm k},r^c_{\bm k}\}_{mn}
                       +r_{mn\bm k}^a\{v^d_{\bm k},r^c_{\bm k}\}_{nm} \notag\\
    &-(v_{mm\bm k}^c+v_{nn\bm k}^c)r_{nm\bm k}^dr_{mn\bm k}^a
    \frac{3\hbar\omega_{mn\bm k}-2\hbar\omega}{\hbar\omega_{mn\bm k}-\hbar\omega^+}\,.
  \end{align}
When simplifying the results of Mahon and Sipe \cite{Mahon2020},  we can find all diagonal terms $\bm \xi_{nn\bm k}$ are canceled out, and only the off-diagonal terms $\bm r_{nm\bm k}$ remain. 
The final simplified expression is found to be the same as $S_r^{dac}(\omega)$ after combining different terms and alternatively using the relation between
$\bm r_{nm\bm k}$ and $\bm v_{nm\bm k}$. 

For crystals with time-reversal symmetry, the matrix elements satisfy \cite{Sipe2000PRB}
  \begin{align}
    \bm r_{nm(-\bm k)} = \bm r_{mn\bm k} \,, \quad 
    \bm v_{nm(-\bm k)} = -\bm v_{mn\bm k}\,.
  \end{align}
  Using this relation, one can show $S_0^{dac}=0$.

\section{Tight-binding model for twisted bilayer graphene\label{app:tb}}

The tight-binding Hamiltonian of a commensurate TBG is adopted from the work by Moon and Koshino \cite{Moon2013}, and slightly modified with the inclusion of the nonequivalent A and B onsite energies.
The Hamiltonian used here is  different from that used by Morell {\it et al.} \cite{Morell2017}, and the details are listed below briefly.

Starting from a AA stacked bilayer graphene, the TBG is constructed by rotating the upper and lower layers by ${\theta}/{2}$ and $-{\theta}/{2}$, respectively, and then translating the layer 2 relative to the layer 1 by an inplane vector $\bm \delta$.
Here, $\bm\delta=\bm 0$ and $\bm\delta=2(\bm a_1+\bm a_2)/3$ represent the rotation center at the atom and at the center of hexagon, and correspond to the $D_3$ and $D_6$ point group, respectively.
Taking the lattice vectors of the unrotated graphene as ${\bm a}_1=a(\sqrt{3}/2, -1/2)$ and ${\bm a}_2=a(\sqrt{3}/2, 1/2)$ with the lattice constant $a\approx 2.46$ {\AA}, the supercell of the commensurate TBG can be described by two integer numbers $(m,n)$ as
\begin{subequations}
\begin{align}
  {\bm A}_1=m{\bm a}^{(\nu)}_1+n{\bm a}^{(\nu)}_2\,,
\end{align}
\begin{align}
  {\bm A}_2=R(\pi/3){\bm A}_1\,,
\end{align}
\end{subequations}
where ${\bm a}_j^{(\nu)}$ is the rotated lattice vector in the $\nu=\pm$
layer. The corresponding twist angle $\theta$ is calculated through
\begin{align}
\cos\theta=\frac{m^2+n^2+4mn}{2(m^2+n^2+mn)}\,.
\end{align}
Therefore, the tight-binding Hamiltonian is given by
\begin{align}
\hat H_0=-\sum_{ij}\left[t(\bm R_i-\bm R_j) + \delta_{ij}\left(s_i\frac{\Delta}{2}+\nu_i\frac{V_g}{2}\right)\right]\hat b_i^\dag \hat b_j  \,.
\label{tb-H}
\end{align}
Here ${\bm R}_i=n_i\bm a_1^{(\nu_i)}+m_i\bm a_2^{(\nu_i)}+\bm\tau_{\alpha_i}^{(\nu_i)}+\bm d^{(\nu_i)}$ 
indicates the atom position with an abbreviated index
$i=(n_im_i\alpha_i\nu_i)$, 
where $\bm\tau_{\alpha_i}^{(\nu_i)}$ 
is the bias of the $\alpha_i=A,B$ atom in the unit cell 
and $\bm d^{(\nu_i)}$ is  the $z$-position of the $\nu_i$ layer; 
$|{\bm R}_i\rangle$ indicates the $2p_z$ orbital at the site $\bm R_i$;
$\Delta$ is the onsite energy with $s_i=1$ for $\alpha_i=A$ and $-1$ for $\alpha_i=B$; 
and $V_g$ gives the layer potential (In experiments, values of $V_g$ in TBG were measured to be up to 1 eV, equating to an electric field of roughly 3 V/nm \cite{Brun2015,Zhang2009}).
The hopping term in the work by Moon and Koshino \cite{Moon2013} is given by 
\begin{align}  
-t(\bm d) =& V_{pp\pi}^0 \exp \left({-\frac{d-a_b}{\delta_0}}\right)\cdot
         \left[1-\left(\frac{\bm d\cdot{\bm e}_z}{d}\right)^2\right]\notag \\
           &+V_{pp\sigma}^0 \exp \left({-\frac{d-d_0}{\delta_0}}\right)\cdot
            \left(\frac{\bm d\cdot{\bm e}_z}{d}\right)^2\,,
\end{align}
with $V^0_{pp\pi} \approx -2.7$ eV, $V^0_{pp\sigma} \approx 0.48$ eV, $a_0=a/\sqrt{3}\approx 1.42$ {\AA},  $d\approx3.35$ {\AA}, and $\delta_0=0.184 ~a\approx0.45$ {\AA}.
In contrast, the hopping term in the work by Morell {\it et al.} \cite{Morell2017} is written as 
\begin{align} 
-t(\bm d)=\gamma\sum_{i,j} \exp \left[{-\beta(|{\bm d}|-d)}\right]\,,
\end{align}
with $\gamma=-0.39$ eV and $\beta=3$. 
The position operator $\hat{\bm r}$ is 
\begin{align}
\hat{\bm r} = \sum_i \bm R_i \hat b_i^\dag \hat b_i\,.
\end{align}
The velocity operator $\hat{\bm v}$ is then 
\begin{align}
\hat{\bm v} &= \frac{1}{i\hbar}[\hat{\bm r}, \hat H_0]\,.
\end{align}

Each supercell of TBG contains $N=4(n^2+m^2+nm)$ carbon atoms, which can be identified by biases $\bm t_{l}$ with an index $l=1,2,\cdots,N$. By matching $\bm R_i= p_i\bm A_1 + q_i \bm A_2 + \bm t_{l_i}$, the bias $\bm t_{l_i}$ can be obtained and the abbreviated index $i$ also stands for integers $(p_iq_il_i)$. After performing the transformation 
\begin{align}
\hat b_{pql} &= \sqrt{\frac{\Omega}{(2\pi)^2}}\int d\bm k e^{i\bm k\cdot(p\bm A_1 + q\bm A_2)}\hat c_{l\bm k} \,,
\end{align}
the Hamiltonian becomes
\begin{align}
\hat H_0 = \sum_{l_1l_2} \int d\bm k H_{l_1l_2\bm k}^0 \hat c_{l_1\bm k}^\dag \hat c_{l_2\bm k}\,,
\end{align}
with 
\begin{align}
H_{l_1l_2\bm k}^0 = \sum_{pq} e^{-i\bm k\cdot(p \bm A_1+q\bm A_2)}t(\bm R_{pql_1}-\bm R_{00l_2}) + \delta_{l_1,l_2} \left[s^{(00l_1)}\frac{\Delta}{2}+\nu^{(00l_1)}\frac{V_g}{2}\right]\,.
\end{align}
It can be diagonalized into 
\begin{align}
\hat H_0 = \sum_s \int d\bm k \varepsilon_{s\bm k}\hat a_{s\bm k}^\dag \hat a_{s\bm k}\,,
\end{align}
through the transformation
\begin{align}
\hat c_{l\bm k} &= \sum_{s} C_{s\bm k;l} \hat a_{s\bm k}\,,
\end{align}
where the wavefunctions $C_{s\bm k;l}$ satisfy the eigenequations
\begin{align}
\sum_{l_2} H_{l_1l_2\bm k}C_{s\bm k;l_2}=\varepsilon_{s\bm k} C_{s\bm k;l_1}\,,
\end{align}
with the band index $s$ and eigenenergy $\varepsilon_{s\bm k}$.
Similar transformation gives the velocity operator as
\begin{align}
\hat{\bm v} &= \int d\bm k \sum_{s_1s_2} \bm v_{s_1s_2\bm k} \hat a_{s_1\bm k}^\dag \hat a_{s_2\bm k}\,.
\end{align}


\bibliography{ref}


\end{document}